\documentclass[aps,prd,twocolumn,showpacs,amsmath,amssymb]{revtex4-1}
\usepackage{amsmath}
\usepackage{graphicx}
\usepackage{subfigure}
\usepackage{epstopdf}
\usepackage{color}
\usepackage{multirow}
\usepackage{setspace}
\usepackage{overpic}
\usepackage{amssymb}
\usepackage[bookmarksnumbered, pdfstartview=FitH,colorlinks,urlcolor=blue, citecolor=blue,linkcolor=blue] {hyperref}
\usepackage{lineno}
\usepackage{bm}
\usepackage{rotating}
\usepackage{siunitx}
\usepackage[utf8]{inputenc}

\let\oldequation\equation
\let\oldendequation\endequation

\renewenvironment{equation}
  {\linenomathNonumbers\oldequation}
  {\oldendequation\endlinenomath}

\begin{document}

%\linenumbers

\title{\bf \boldmath{
Improved measurement of the branching fraction of $D_s^+\to\mu^+\nu_\mu$}{}
}

\author{
M.~Ablikim$^{1}$, M.~N.~Achasov$^{13,b}$, P.~Adlarson$^{75}$, X.~C.~Ai$^{81}$, R.~Aliberti$^{36}$, A.~Amoroso$^{74A,74C}$, M.~R.~An$^{40}$, Q.~An$^{71,58}$, Y.~Bai$^{57}$, O.~Bakina$^{37}$
% ...
\\
\vspace{0.2cm}
(BESIII Collaboration)\\
\vspace{0.2cm} {\it
$^{1}$ Institute of High Energy Physics, Beijing 100049, People's Republic of China\\
% ...
$^{13}$ G.I. Budker Institute of Nuclear Physics SB RAS (BINP), Novosibirsk 630090, Russia\\
% ...
$^{36}$ Johannes Gutenberg University of Mainz, Johann-Joachim-Becher-Weg 45, D-55099 Mainz, Germany\\
$^{37}$ Joint Institute for Nuclear Research, 141980 Dubna, Moscow region, Russia\\
% ...
$^{40}$ Liaoning Normal University, Dalian 116029, People's Republic of China\\
% ...
$^{57}$ Southeast University, Nanjing 211100, People's Republic of China\\
$^{58}$ State Key Laboratory of Particle Detection and Electronics, Beijing 100049, Hefei 230026, People's Republic of China\\
% ...
$^{71}$ University of Science and Technology of China, Hefei 230026, People's Republic of China\\
% ...
$^{74}$ University of Turin and INFN, (A)University of Turin, I-10125, Turin, Italy; (B)University of Eastern Piedmont, I-15121, Alessandria, Italy; (C)INFN, I-10125, Turin, Italy\\
$^{75}$ Uppsala University, Box 516, SE-75120 Uppsala, Sweden\\
% ...
$^{81}$ Zhengzhou University, Zhengzhou 450001, People's Republic of China\\
\vspace{0.2cm}
% ...
$^{b}$ Also at the Novosibirsk State University, Novosibirsk, 630090, Russia\\
% ...
}
%% ends here %%
}

\begin{abstract}
Using $e^+e^-$ collision data with an integrated luminosity of $7.33~\mathrm{fb}^{-1}$ collected at center-of-mass energies between 4.128 and 4.226 GeV with the BESIII detector operating at the BEPCII collider, the branching fraction of the leptonic decay $D_s^+\to\mu^+\nu_\mu$ is measured
to be $(0.5294\pm0.0108_{\rm stat}\pm0.0085_{\rm syst})$\%.
Based on this, the product of the $D_s^+$ decay constant $f_{D_s^+}$ and the magnitude of the $c\to s$ quark mixing matrix element $|V_{cs}|$ is determined to be $f_{D_s^+}|V_{cs}|=241.8\pm2.5_{\rm stat}\pm2.2_{\rm syst}~\mathrm{MeV}$.
Using the value of $|V_{cs}|$ given by the global standard model fit, $f_{D_s^+}$ is found to be $248.4\pm2.5_{\rm stat}\pm2.2_{\rm syst}$\,MeV. Alternatively, using the value of $f_{D_s^+}$ from a recent lattice quantum chromodynamics calculation, $|V_{cs}|$ is determined to be $0.968\pm0.010_{\rm stat}\pm0.009_{\rm syst}$.
\end{abstract}

\maketitle
\section{Introduction}
Experimental studies of the leptonic decay $D^+_s\to \ell^+\nu_\ell$~($\ell=e$, $\mu$ or $\tau$) are important to explore both the strong and weak interactions in the charm quark sector. In the standard model~(SM), the $D^+_s\to \ell^+\nu_\ell$ decay partial width is given by~\cite{decayrate}
\begin{equation}
\Gamma_{D^+_{s}\to\ell^+\nu_\ell}=\frac{G_F^2}{8\pi}|V_{cs}|^2
f^2_{D^+_{s}}
m_\ell^2 m_{D^+_{s}} \left (1-\frac{m_\ell^2}{m_{D^+_{s}}^2} \right )^2,
\end{equation}
where
$G_F$ is the Fermi coupling constant,
$m_\ell$ is the lepton mass,
$m_{D^+_{s}}$ is the $D^+_{s}$ mass,
$f_{D^+_{s}}$ is the $D^+_{s}$ decay constant, and
$|V_{cs}|$ is the magnitude of the $c\to s$
Cabibbo-Kobayashi-Maskawa~(CKM) matrix element.

In recent years, many studies of $D^+_s\to \ell^+\nu_\ell$ have been performed by the CLEO~\cite{cleo2009,cleo2009a,cleo2009b}, BaBar~\cite{babar2010}, Belle~\cite{belle2013}, and
BESIII~\cite{bes2016,besiii2,Hajime,bes32,bes33,bes34,bes35} experiments.
The experimental precision of $f_{D^+_s}$, however, is still worse than the
lattice quantum chromodynamics~(LQCD)
calculation~\cite{FLab2018,Ke:2023qzc}.
Intensive experimental studies of $D^+_s\to \ell^+\nu_\ell$ can determine $f_{D^+_{s}}$ and $|V_{cs}|$ with improved precision, and can thereby calibrate various theoretical calculations of $f_{D^+_{s}}$~\cite{FLab2018,LQCD,etm2015,ukqcd2017,milc2012,hpqcd2010,hpqcd2012,etm2012,chen2014,pacs2011,becirevic2013}
and test the unitarity of the CKM matrix.

In addition, scrutinizing lepton flavor universality~(LFU) violation in $D^+_s\to\ell^+\nu_\ell$ decays offers an important test of the SM~\cite{babar_1,babar_2,lhcb_1,lhcb_2,lhcb_3,belle_RD1,belle_RD2,belle_RD3,HFLAV}. In the SM, the ratio of the branching fraction (BF) of $D^+_s\to \tau^+\nu_\tau$
over that of $D^+_s\to \mu^+\nu_\mu$ is predicted to be 9.75 with a
negligible uncertainty, hence any observed deviation
from this value would means a sign of new physics beyond the SM.
Some hints of LFU violation in semileptonic $B$ decays were reported at BaBar, LHCb, and Belle~\cite{babar_1,babar_2,lhcb_1,lhcb_2,lhcb_3,belle_RD1,belle_RD2,belle_RD3}.
The measured BF ratios $\mathcal{R}_{D^{(*)}}^{\tau/\ell}=\mathcal{B}_{B\to \bar D^{(*)}\tau^+\nu_{\tau}}/\mathcal{B}_{B\to \bar D^{(*)}\ell^+\nu_{\ell}}$ deviate from the SM predictions by about $3.3\sigma$~\cite{HFLAV}.
A test of LFU with $D^+_s\to \ell^+\nu_\ell$ decays may shed light on this tension.

In this paper, we report an improved measurement of
the BF of the  $D_s^+\to\mu^+\nu_\mu$ decay
by analyzing 7.33\,fb$^{-1}$ of $e^+e^-$
collision data collected with the BESIII detector
at center-of-mass energies of $E_{\rm cm} = 4.128, 4.156, 4.178, 4.189, 4.199, 4.209, 4219,$ and $4.226$~GeV. Charge-conjugate ($c.c.$) modes are always implied in the text. Benefiting from a larger data sample, more tag modes, and measurement from muon identifier modules, the results obtained in this work supersede the previous BESIII measurement with the muon identifier modules using data at $E_{\rm cm}=4.178$~GeV~\cite{besiii2} and the measurement without the muon identifier modules using data at $E_{\rm cm}$ between 4.178 and 4.226~GeV~\cite{Hajime}.

\section{BESIII detector and Monte Carlo simulation}

The BESIII detector is a magnetic
spectrometer~\cite{BESIII} located at the Beijing Electron
Positron Collider (BEPCII)~\cite{Yu:IPAC2016-TUYA01}. The
cylindrical core of the BESIII detector consists of a helium-based
 multilayer drift chamber (MDC), a plastic scintillator time-of-flight
system (TOF), and a CsI (Tl) electromagnetic calorimeter (EMC),
which are all enclosed in a superconducting solenoidal magnet
providing a 1.0~T magnetic field. The solenoid is supported by an
octagonal flux-return yoke with resistive plate counter muon
identifier modules interleaved with steel. The solid angle coverage for detecting charged particles is 93\% over $4\pi$. The
charged-particle momentum resolution at $1~{\rm GeV}/c$ is
$0.5\%$, and the resolution of the specific ionization energy loss~(d$E$/d$x$) is $6\%$ for the electrons
from Bhabha scattering. The EMC measures photon energies with a
resolution of $2.5\%$ ($5\%$) at $1$~GeV in the barrel (end cap)
region. The time resolution of the TOF barrel part is 68~ps, while
that of the end cap part is 110~ps. The end cap TOF
system was upgraded in 2015 using multi-gap resistive plate chamber technology, providing
a time resolution of 60 ps~\cite{60ps1,60ps2}.
Approximately 83\%
of the data used here was collected after this upgrade, and the corresponding luminosities~\cite{luminosities,luminosities2} at each energy are given in Table~\ref{tab:mbc}.
More details about the design and performance of the BESIII detector are given in Ref.~\cite{BESIII}.

Simulated samples produced with {\sc geant4}-based~\cite{geant4}
Monte Carlo (MC) software, which includes the geometric description of
the BESIII detector and the detector response, are used to determine
the detection efficiency and to estimate background contributions. The
simulations include the beam energy spread and initial state radiation
in the $e^+e^-$ annihilations modeled with the generator {\sc
  kkmc}~\cite{kkmc,kkmc2}. An inclusive MC sample with an equivalent luminosity of 40 times that of the data is produced at  center-of-mass energies between 4.128 and 4.226 GeV. It includes open-charm processes, initial state radiation (ISR) production of $\psi(3770)$, ISR production of $\psi(3686)$, ISR production of $J/\psi$, $q\bar q$ $(q=u, d, s)$ continuum processes, Bhabha scattering, $\mu^+\mu^-$, $\tau^+\tau^-$, and $\gamma\gamma$ events. In the simulation, the production of open-charm processes directly produced via $e^+e^-$ annihilations are modeled with the generator {\sc conexc}~\cite{conexc}, and their subsequent decays are modeled by {\sc evtgen}~\cite{evtgen,evtgen2} with known BFs from the Particle Data Group (PDG)~\cite{PDG2022}. The input cross
section of $e^+e^-\to D_s^{\pm}D_s^{*\mp}$ is based on the cross section measurement in the energy range from threshold to 4.7 GeV. The ISR production of vector charmonium(-like) states and the continuum processes are incorporated in  {\sc
  kkmc}~\cite{kkmc,kkmc2}. The remaining unknown decays of the
charmonium states are modeled by {\sc
  lundcharm}~\cite{lundcharm,lundcharm2}. Final state radiation is
incorporated using {\sc photos}~\cite{photos}.

\section{Method}
\label{sec:method}

At $E_{\rm cm}$ from 4.128 to 4.226 GeV, the $D^{\pm}_s$ mesons are produced mainly through the process $e^+e^-\to D^{*\pm}_sD_s^{\mp}\to\gamma(\pi^0)D^{\pm}_sD_s^{\mp}$.
A single-tag (ST) sample is selected by fully reconstructing a $D_s^-$ meson in one of several `tag' modes.
Those in which the ST $D_s^-$, the transition $\gamma(\pi^0)$ of the $D_s^{*+}$ decay and the signal $D_s^+$ decay of interest are simultaneously reconstructed are called double-tag (DT) sample.
Then the BF of $D^+_s\to \mu^+\nu_\mu$ is determined by
\begin{equation}
\mathcal B_{D_s^+\to\mu^+\nu_\mu}=\frac{N_{\rm DT}}{N_{\rm ST} \cdot \epsilon_{\gamma(\pi^0)\mu^+\nu_\mu}},
\label{eq1}
\end{equation}
where $N_{\rm DT}$ is the DT yield in data, $N_{\rm ST}=\Sigma_i N_{\rm ST}^{i}$ is the total ST yield in data summing over tag mode $i$, $\epsilon_{\gamma(\pi^0)\mu^+\nu_\mu}=\Sigma_i(N_{\rm ST}^{i}/N_{\rm ST}) \cdot (\epsilon^{i}_{\rm DT}/\epsilon^{i}_{\rm ST})$ is the effective signal efficiency of detecting $D^+_s\to \mu^+\nu_\mu$ in the presence of the ST $D^-_s$ candidate, averaged by the ST yields in data, and $\epsilon^{i}_{\rm DT}$ and $\epsilon^{i}_{\rm ST}$ are the detection efficiencies of the DT and ST candidates, respectively.

\section{Single Tag Selection}

The ST $D_s^-$ mesons are reconstructed using 16 hadronic $D_s^-$ decay modes:
$K^+K^-\pi^-$,
$K^+K^-\pi^-\pi^0$,
$\pi^+\pi^-\pi^-$,
$K_S^0K^-$,
$K_S^0K^-\pi^0$,
$K^-\pi^+\pi^-$,
$K_S^0K_S^0\pi^-$,
$K_S^0K^+\pi^-\pi^-$,
$K_S^0K^-\pi^+\pi^-$,
$\eta_{\gamma\gamma}\pi^-$,
$\eta_{\pi^+\pi^-\pi^0}\pi^-$,
$\eta^\prime_{\pi^+\pi^-\eta_{\gamma\gamma}}\pi^-$,
$\eta^\prime_{\gamma\rho^0}\pi^-$,
$\eta_{\gamma\gamma}\rho^-$,
$\eta_{\pi^+\pi^-\pi^0}\rho^-$, and
$\eta_{\gamma\gamma}\pi^+\pi^-\pi^-$,
where the subscripts on the $\eta(\eta^{\prime})$ represent the decay modes used to reconstruct the $\eta(\eta^{\prime})$.

All charged tracks except for those from $K_S^0$ decays must originate from the interaction
point~(IP) with a distance of closest approach less than 1 cm in the transverse plane
and less than 10 cm along the $z$ axis. The polar angle ($\theta$) is required to be within the MDC
acceptance $|\rm{cos\theta}|<0.93$, where $\theta$ is defined with
respect to the symmetry axis of the MDC taken as the $z$ axis. Measurements of the $\text{d}E/\text{d}x$ in the MDC and the flight time in the TOF are combined for particle identification~(PID) by calculating
confidence levels for the pion and kaon hypotheses ($CL_\pi$,~$CL_K$).
Kaon~(pion) candidates are required to satisfy $CL_{K(\pi)}>CL_{\pi(K)}$.

To select $K_S^0$ candidates, pairs of oppositely-charged tracks with distance
of closest approach to the IP less than 20 cm along the $z$ axis are assigned as
$\pi^+\pi^-$ without PID requirements.
These $\pi^+\pi^-$ combinations are required to
have an invariant mass within $\pm12$\,MeV of the nominal $K_S^0$ mass~\cite{PDG2022} and
a decay length greater than twice the vertex resolution away from the IP.

The $\pi^0$ and $\eta$ mesons are reconstructed via their decays to $\gamma\gamma$. Here, the $\gamma$ candidates are identified using isolated showers in the EMC. The deposited energy of each shower must be greater than 25 (50) MeV in the EMC barrel
(end-cap) region~\cite{BESIII}.
To suppress electronic noise and showers
unrelated to the event, the difference between the EMC time and
the event start time is required to be within $[0, 700]$ ns~\cite{BESIII}.
The opening angle between a shower and
the nearest charged track has to be greater than $10^{\circ}$.
The $\gamma\gamma$ combinations with invariant masses $M_{\gamma\gamma}\in(0.115,\,0.150)$
and $(0.500,\,0.570)$\,GeV$/c^{2}$ are regarded as $\pi^0$ and $\eta$ candidates, respectively.
A kinematic fit is further performed to
constrain $M_{\gamma\gamma}$ to the $\pi^{0}$ or $\eta$
nominal mass~\cite{PDG2022}.

The $\eta$ candidates for the $\eta\pi^-$
ST channel are also reconstructed
via $\pi^0\pi^+\pi^-$ final states with an invariant mass
within $(0.53,\,0.57)~\mathrm{GeV}/c^2$.
The $\eta^\prime$ mesons are reconstructed via two decay modes, $\eta\pi^+\pi^-$ and $\gamma\rho^0$,
whose invariant masses are required to be within
$(0.946,\,0.970)$ and $(0.940,\,0.976)~\mathrm{GeV}/c^2$, respectively.
In addition, the minimum energy
of the $\gamma$ from $\eta^\prime\to\gamma\rho^0$ decays must be greater than 0.1\,GeV.
The $\rho^0$ and $\rho^+$ mesons are reconstructed from $\pi^+\pi^-$ and $\pi^+\pi^0$
combinations whose invariant masses are required to be within $(0.67,\,0.87)~\mathrm{GeV}/c^2$.

The momentum of any pion not originating from a $K_S^0$, $\eta$, or $\eta^\prime$ decay is required to be greater than 0.1\,GeV/$c$
to reject transition pions from $D^*$ decays.
When selecting $\pi^+\pi^-\pi^-$ and $K^-\pi^+\pi^-$ combinations, peaking backgrounds from
$K^0_S\pi^-$ and $K_S^0K^-$ components are rejected by requiring that the invariant mass of any $\pi^+\pi^-$ combination
satisfy $|M_{\pi^+\pi^-}-M_{K_S^0}|>0.030~$~GeV/$c^2$~\cite{PDG2022}.

\begin{figure*}[htbp]
  \begin{center}
\includegraphics[width=0.8\textwidth] {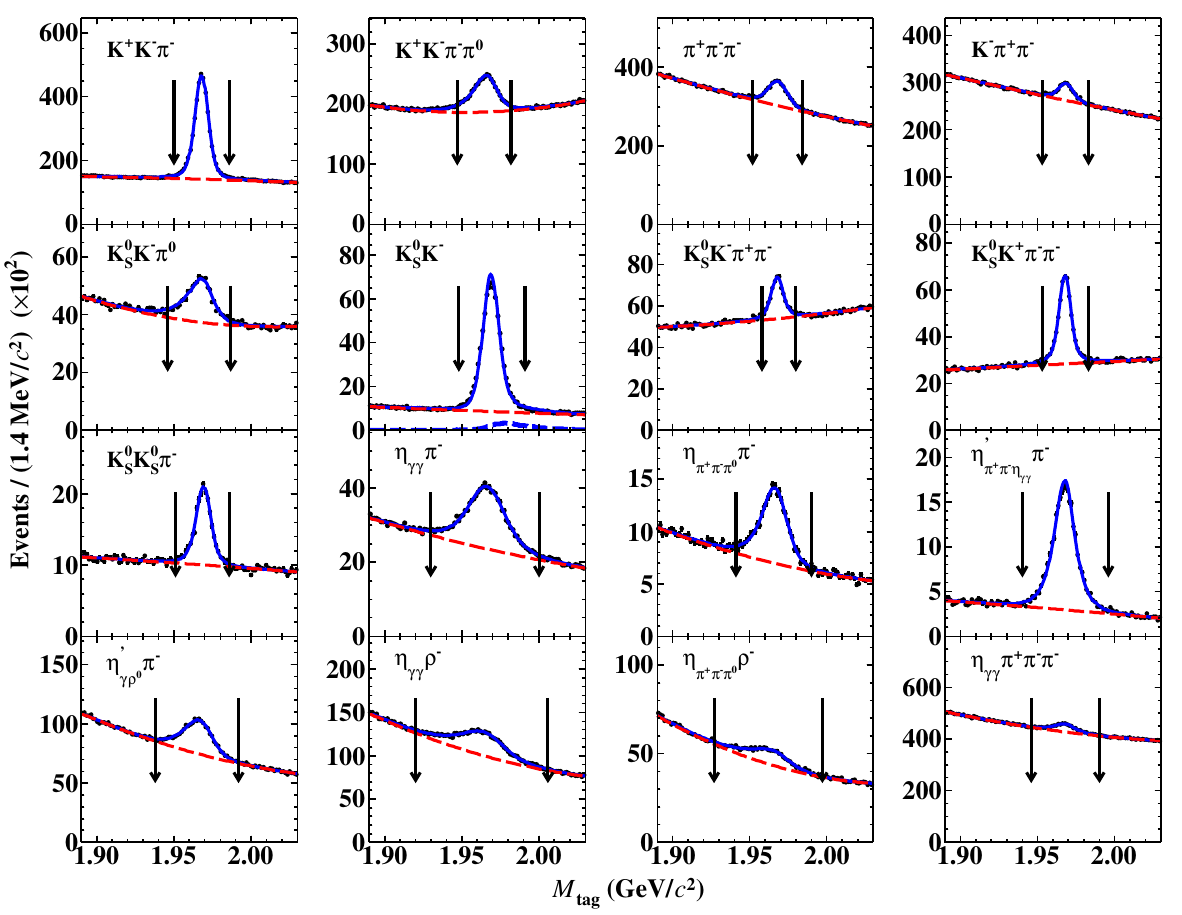}
  \caption{ Fits to the $M_{\rm tag}$ distributions of ST $D^-_s$ candidates selected from data at all energy points, where the points with error bars are data, the solid curve shows the best fit, and the red dashed line shows the shape of the combinatorial backgrounds.
  In the fit to the $M_{\rm tag}$ distribution for $D_s^-\to K_S^0K^-$, the blue dashed curve shows the shape of $D^-\to K_S^0\pi^-$. }
  \label{fig:stfit}
  \end{center}
\end{figure*}

The beam-constrained mass of the ST $D_s^-$
candidate
\begin{equation}
M_{\rm BC}\equiv\sqrt{(E_{\rm cm}/2)^2/c^4-|\vec{p}_{D_s^-}|^2/c^2}
\end{equation}
is used to suppress the non-$D_s^+D^{*-}_s$ events,
where
$\vec{p}_{D_s^-}$ is the momentum of the ST $D_s^-$ candidate. The $M_{\rm BC}$ of the direct $D^-_s$ candidate form a peak around 2.04 GeV/$c^2$ with a resolution of (1.9-2.8) MeV, while it ranges in $(2.01, 2.07)$ GeV/$c^2$ for the in-direct $D^-_s$ candidate.
The requirements on $M_{\rm BC}$ for the eight center-of-mass energies are listed in Table~\ref{tab:mbc}, which retain the $D_s^-$ mesons
from both the $e^+e^-$ annihilation
and the $D_s^{*-}$ decay. The efficiency loss due to the $M_{\rm BC}$ selection is not more than 3\% for each energy point. In each event, we only keep
the candidate with the $D_s^-$ recoil mass
\begin{equation}
M_{\rm rec} \equiv \sqrt{ \left (E_{\rm cm} - \sqrt{|\vec p_{D^-_s}|^2c^2+m^2_{D^-_s}c^4} \right )^2/c^4
-|\vec p_{D^-_s}|^2/c^2}
\end{equation}
closest to the nominal $D_s^{*+}$ mass~\cite{PDG2022} per tag mode per charge. The probability of
the best candidate selection for individual tag modes ranges in (82-99)\%.
Figure~\ref{fig:stfit} shows the invariant mass ($M_{\rm tag}$) spectra of the accepted ST
candidates with all datasets.

\begin{table}[htbp]
\centering
\caption{The integrated luminosities and requirements on $M_{\rm BC}$ for the ST candidates at various energy points.}
\begin{tabular}{cSc}
\hline
\hline
$E_{\rm cm}$ (GeV)  &{Luminosity (pb$^{-1}$)}& $M_{\rm BC}$ (GeV/$c^2$) \\\hline
4.128               &401.5    & $(2.010,2.061)$                        \\
4.157               &408.7    & $(2.010,2.070)$                        \\
4.178               &3189.0   & $(2.010,2.073)$                        \\
4.189               &569.8    & $(2.010,2.076)$                        \\
4.199               &526.0    & $(2.010,2.079)$                        \\
4.209               &571.7    & $(2.010,2.082)$                       \\
4.219               &568.7    & $(2.010,2.085)$                        \\
4.226               &1091.7   & $(2.010,2.088)$                        \\
\hline
\hline
\end{tabular}
\label{tab:mbc}
\end{table}

At each energy point, the ST yield for each tag mode is obtained by a fit to the corresponding $M_{\rm tag}$ spectrum.
The signal is described by the MC-simulated shape convolved with a Gaussian function
representing the resolution difference between data and MC simulation.
For the tag mode $D^-_s\to K_S^0K^-$,
the peaking background from $D^-\to K^0_S\pi^-$ is
described by the MC-simulated shape smeared with the same Gaussian function
as used in the signal shape with the magnitude free in the fit.
The non-peaking background is modeled by a second- or third-order Chebychev polynomial
function, which is validated with the inclusive MC sample.
The fit results of $M_{\rm tag}$ summed over
all energy points are shown in Fig.~\ref{fig:stfit}.
The events in the signal regions
are kept for further analysis. The fraction of the $e^+e^-\to\gamma(\pi^0)D^{\pm}_sD_s^{\mp}$
process is about (0.7-1.1)\% in the fitted yield of ST $D_s^-$ mesons for each tag mode, and has been subtracted away from individual fitted ST yield.
As an example, the ST yields in data
and the ST efficiencies at 4.178 GeV are shown in Table~\ref{tab:styields}.
The total ST yields at the different energy points are summarized in Table~\ref{tab:sigeff}.

\begin{table*}[htbp]
\caption{The $M_{\rm tag}$ requirements, ST yields in data ($N_{\rm ST}^{i}$), ST efficiencies ($\epsilon_{\rm ST}^{i}$), and effective signal efficiencies with transition $\gamma(\pi^0)$ matched ($\epsilon^{i,\rm matched}_{\gamma(\pi^0) \mu^+\nu_\mu}$) of various tag modes obtained from data at $E_{\rm cm}=4.178$ GeV.
The uncertainties are statistical only. The $\epsilon^{i,\rm matched}_{\gamma(\pi^0) \mu^+\nu_\mu}$ varies within 37\% for different tag modes, which are mainly caused by the significantly different signal environments for some tag modes containing low momentum photon and pions in the signal and inclusive MC samples. The ST efficiency of $D_s^-\to K_S^0K^-\pi^+\pi^-$ is $\sim$12\% lower than that of $D_s^-\to K_S^0K^-\pi^+\pi^-$, due to more low momentum charged particles from different subresonances.}
\label{tab:styields}
\centering
\begin{tabular}{lc r@{\,}c@{\,}l r@{\,}c@{\,}l r@{\,}c@{\,}l}
\hline\hline
Tag mode & $M_{\rm tag}$ (GeV/$c^2$) &\multicolumn{3}{c}{$N_{\rm ST}^{i}$} &\multicolumn{3}{c}{$\epsilon_{\rm ST}^{i}$ (\%)} &\multicolumn{3}{c}{$\epsilon^{i,\rm matched}_{\gamma(\pi^0) \mu^+\nu_\mu}$ (\%)}\\
\hline
$K^+K^-\pi^-$                           &$(1.950,1.986)$& $137317$&$\pm$&$608$  &$40.92$&$\pm$&$0.02$&$49.25$&$\pm$&$ 0.21$\\
$K^+K^-\pi^-\pi^0$                      &$(1.947,1.982)$& $42119$&$\pm$&$851$  &$11.77$&$\pm$&$0.01$&$58.33$&$\pm$&$ 0.74$\\
$\pi^-\pi^+\pi^-$                       &$(1.952,1.984)$& $36497$&$\pm$&$873$  &$52.13$&$\pm$&$0.05$&$51.16$&$\pm$&$ 0.19$\\
$K_S^0 K^-$                             &$(1.948,1.991)$& $30956$&$\pm$&$261$  &$47.63$&$\pm$&$0.05$&$50.32$&$\pm$&$ 0.23$\\
$K_S^0 K^-\pi^0$                        &$(1.946,1.987)$& $11182$&$\pm$&$449$  &$17.01$&$\pm$&$0.04$&$58.42$&$\pm$&$ 0.74$\\
$K^-\pi^+\pi^-$                         &$(1.953,1.983)$& $16514$&$\pm$&$632$  &$45.42$&$\pm$&$0.07$&$51.44$&$\pm$&$ 0.41$\\
$K_S^0 K_S^0 \pi^-$                     &$(1.951,1.986)$& $5088$&$\pm$&$149$   &$22.82$&$\pm$&$0.07$&$51.97$&$\pm$&$ 0.91$\\
$K_S^0 K^+\pi^-\pi^-$                   &$(1.953,1.983)$& $14855$&$\pm$&$235$  &$21.12$&$\pm$&$0.04$&$50.72$&$\pm$&$ 0.79$\\
$K_S^0 K^-\pi^+\pi^-$                   &$(1.958,1.980)$& $7621$&$\pm$&$270$   &$18.51$&$\pm$&$0.05$&$52.15$&$\pm$&$ 0.43$\\
$\eta\pi^-$                             &$(1.930,2.000)$& $19239$&$\pm$&$468$  &$48.79$&$\pm$&$0.06$&$54.12$&$\pm$&$ 0.20$\\
$\eta_{\pi^+\pi^-\pi^0}\pi^-$           &$(1.941,1.990)$& $5693$&$\pm$&$201$   &$23.49$&$\pm$&$0.07$&$55.51$&$\pm$&$ 0.34$\\
$\eta^\prime_{\eta\pi^+\pi^-}\pi^-$     &$(1.940,1.996)$& $9730$&$\pm$&$140$   &$25.26$&$\pm$&$0.05$&$55.12$&$\pm$&$ 0.31$\\
$\eta^\prime_{\gamma\rho}\pi^-$         &$(1.938,1.992)$& $24698$&$\pm$&$656$  &$32.53$&$\pm$&$0.04$&$54.10$&$\pm$&$ 0.26$\\
$\eta\rho^-$                            &$(1.920,2.006)$& $39670$&$\pm$&$1673$ &$19.88$&$\pm$&$0.02$&$66.92$&$\pm$&$ 0.37$\\
$\eta_{\pi^+\pi^-\pi^0}\rho^-$          &$(1.927,1.997)$& $10504$&$\pm$&$928$  &$ 9.23$&$\pm$&$0.02$&$67.48$&$\pm$&$ 0.57$\\
$\eta_{\gamma\gamma}\pi^+\pi^-\pi^-$    &$(1.946,1.990)$& $23417$&$\pm$&$1232$ &$24.84$&$\pm$&$0.04$&$63.74$&$\pm$&$ 0.32$\\
\hline
\hline
\end{tabular}
\end{table*}

\begin{table}[htbp]\centering
  \caption{The ST yields in data ($N_{\rm ST}$), the weighted signal efficiencies with transition $\gamma(\pi^0)$ matched ($\epsilon^{\rm matched}_{\gamma(\pi^0) \mu^+\nu_\mu}$), and signal efficiencies unmatched within $20^\circ$ ($\epsilon^{\rm unmatched}_{\gamma(\pi^0) \mu^+\nu_\mu}$) for various energy points. The uncertainties are statistical only. The weighted signal efficiencies monotonically decrease with center-of-mass energy, mainly due to different ISR and final state effects as well as correction factors for muon identification.}
  \label{tab:sigeff}
  \begin{tabular}{cr@{\,}c@{\,}l r@{\,}c@{\,}l r@{\,}c@{\,}l}
  \hline
  \hline
  $E_{\rm cm}$ (GeV) & \multicolumn{3}{c}{$N_{\rm ST}$}  &\multicolumn{3}{c}{$\epsilon^{\rm matched}_{\gamma(\pi^0) \mu^+\nu_\mu}$} (\%)  & \multicolumn{3}{c}{$\epsilon^{\rm unmatched}_{\gamma(\pi^0) \mu^+\nu_\mu}$ (\%)}\\ \hline
  4.128 &$38203$&$\pm$&$1794$   &$57.48$&$\pm$&$0.14$  &$5.91$&$\pm$&$0.05$\\
  4.157 &$55280$&$\pm$&$971$    &$56.26$&$\pm$&$0.13$  &$5.96$&$\pm$&$0.05$\\
  4.178 &$435100$&$\pm$&$2926$  &$54.35$&$\pm$&$0.12$  &$5.64$&$\pm$&$0.05$\\
  4.189 &$72852$&$\pm$&$1030$   &$52.76$&$\pm$&$0.12$  &$5.67$&$\pm$&$0.05$\\
  4.199 &$68054$&$\pm$&$1136$   &$53.31$&$\pm$&$0.12$  &$5.94$&$\pm$&$0.05$\\
  4.209 &$68608$&$\pm$&$1281$   &$51.12$&$\pm$&$0.12$  &$5.70$&$\pm$&$0.05$\\
  4.219 &$58151$&$\pm$&$1159$   &$49.47$&$\pm$&$0.12$  &$5.41$&$\pm$&$0.04$\\
  4.226 &$91218$&$\pm$&$1739$   &$49.66$&$\pm$&$0.12$  &$5.62$&$\pm$&$0.05$\\
  \hline\hline
  \end{tabular}
\end{table}

\section{Double Tag Selection}

We select candidates for the transition $\gamma(\pi^0)$ from the $D_s^{*+}$ decay among the unused particles recoiling against the ST $D_s^-$, by using the kinematic variable
\begin{equation}
\Delta E \equiv E_{\rm cm}-E_{\rm tag}-\sqrt{|-\vec{p}_{\rm tag}-\vec{p}_{\gamma(\pi^0)}|^2/c^4+m_{D_s^+}^2}-E_{\gamma(\pi^0)},
\end{equation}
where $E_i$ and $\vec p_i$, with $i=[\gamma(\pi^0)$ or tag], denote
the energy and momentum of particle $i$.
We loop over all remaining $\gamma$ or $\pi^0$ candidates
and choose the one giving a minimum $|\Delta E|$.
The events with $\Delta E\in(-0.05,\,0.10)$\,GeV are accepted.

\begin{figure}[htbp]
  \centering
  \includegraphics[width=0.48\textwidth]{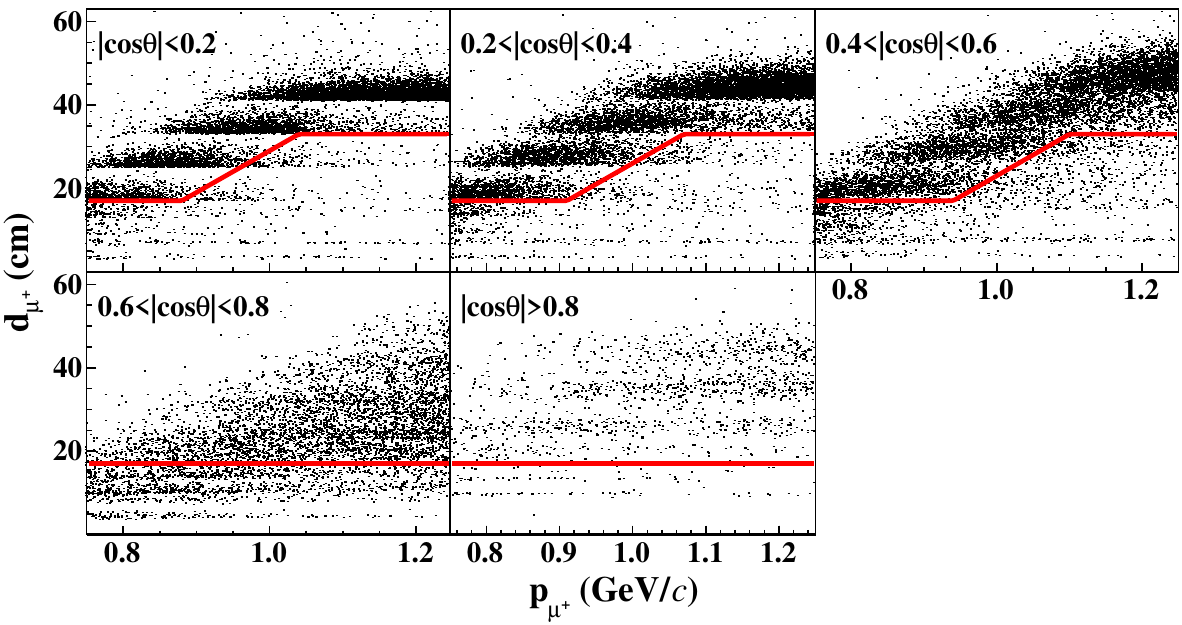}
\caption{The $d_{\mu^+}$ vs. $p_{\mu^+}$ of muon candidates of the $e^+e^-\to \gamma \mu^+\mu^-$ processes in different $|\cos\theta |$ regions from data.}
  \label{fig:muid_cos}
\end{figure}

In the presence of ST $D_s^-$ and transition $\gamma(\pi^0)$,
the $D_s^+\to\mu^+\nu_\mu$ candidates are selected using the
remaining neutral and charged tracks. The muon candidate is required to have an opposite charge
to the ST $D^-_s$ meson and
deposited energy in the EMC within $(0.0,\,0.3)$\,GeV. To separate muons from hadrons,  fulfill requirements on the muon hit depth ($d_{\mu^+}$) in the muon counter with dependence of $p_{\mu^+}$ and flight direction $\cos\theta$ in the muon identifier modules.
To consider the $p_{\mu^+}$ and $d_{\mu^+}$ dependence, we examine the distributions of $d_{\mu^+}$ versus $p_{\mu^+}$ using $e^+e^-\to(\gamma)\mu^+\mu^-$ candidates selected from data, as shown in Fig.~\ref{fig:muid_cos}.
The $|\cos\theta_{\mu^+}|$ and $p_{\mu^+}$ dependent requirements on $d_{\mu^+}$ are shown in
Table~\ref{tab:muonid}.

\begin{table}[hbtp]
  \begin{center}
  \caption{The $\cos\theta_{\mu^+}$ and $p_{\mu^+}$ dependent requirements
    of $d_{\mu^+}$ for muon candidates. To compute the $d_{\mu^+}$ requirements, the momentum
  $p_{\mu^+}$ is taken in units of GeV/$c$.}
  \begin{tabular}{c|c|c} \hline\hline
  $|\cos\theta_{\mu^+}|$&$p_{\mu^+}$ (GeV/$c$)& $d_{\mu^+}$ (cm) \\ \hline
           &$p_{\mu^+}\le0.88$      &$>17.0$        \\
  $$(0.00,0.20)$$&$0.88<p_{\mu^+}<1.04$&$>100.0\times p_{\mu^+}-71.0$ \\
           &$p_{\mu^+}\ge1.04$      &$>33.0$ \\ \hline
           &$p_{\mu^+}\le0.91$      &$>17.0$        \\
  $$(0.20,0.40)$$&$0.91<p_{\mu^+}<1.07$&$>100.0\times p_{\mu^+}-74.0$ \\
           &$p_{\mu^+}\ge1.07$      &$>33.0$ \\ \hline
           &$p_{\mu^+}\le0.94$      &$>17.0$        \\
  $$(0.40,0.60)$$&$0.94<p_{\mu^+}<1.10$&$>100.0\times p_{\mu^+}-77.0$ \\
           &$p_{\mu^+}\ge1.10$      &$>33.0$ \\ \hline
  $$(0.60,0.80)$$&                  &$>17.0$ \\ \hline
  $$(0.80,0.93)$$&                  &$>17.0$ \\ \hline\hline
  \end{tabular}
  \label{tab:muonid}
  \end{center}
\end{table}

To suppress backgrounds with extra photon(s),
the maximum energy of the unused showers in the DT
selection ($E^{\mathrm{extra}~\gamma}_{\rm max}$) is required to be less than 0.3\,GeV.
No any additional charged track is allowed in the event. The yield of signal events is determined by a fit to the distribution of the kinematic variable
\begin{align}
M_{\rm miss}^2&\equiv E_{\nu}^2/c^4-|\vec{p}_{\nu}|^2/c^2.
\end{align}
Here $E_{\nu}\equiv E_{\rm cm}-E_{\rm tag}-E_{\gamma(\pi^0)}-E_{\mu}$ and $\vec p_{\nu}\equiv-\vec{p}_{\rm tag}-\vec{p}_{\gamma(\pi^0)}-\vec{p}_{\mu}$, where $E_{\mu}$ and $\vec{p}_{\mu}$ denote
the energy and momentum of the muon, respectively. $M_{\rm miss}^2$ is the missing mass square of the undetected neutrino.
To improve the $M_{\rm miss}^2$ resolution,
the candidate tracks plus the missing neutrino
are subjected to a 4-constraint kinematic fit requiring energy and momentum conservation.
In addition, the invariant masses of the two $D_s$ mesons are constrained to the known $D_s$ mass,
and the invariant mass of the $D_s^-\gamma(\pi^0)$ or $D_s^+\gamma(\pi^0)$ combination is
constrained to the nominal $D_s^*$ mass.
The combination with the minimum $\chi^2$ is kept.
Figure~\ref{fig:mm2ft} shows the $M_{\rm miss}^2$
distribution for the accepted DT candidate events in data.

The efficiencies of the DT reconstruction are determined with the signal MC samples.
Dividing them by the ST efficiencies determined with the inclusive MC sample yields the corresponding
efficiencies of the $\gamma(\pi^0)\mu^+\nu_\mu$ reconstruction.
The efficiency averaged over all tag modes
is then determined from
\begin{equation}
\epsilon_{\gamma(\pi^0)\mu^+\nu_\mu}=f_{\mu\,\rm PID}^{\rm cor}f_{\rm tag~bias}^{\rm cor}\sum_{i}(N_{\rm ST}^i\epsilon_{\rm DT}^i)/(N_{\rm ST}^{\rm tot}\epsilon_{\rm ST}^i).
\end{equation}
In the above equation, the correction factor $f_{\mu\,\rm PID}^{\rm cor}$ accounts for the differences of $\mu^+$ identification efficiencies between data and MC simulation. It is non-negligible mainly due to the imperfect simulation of $d_{\mu^+}$~\cite{besiii2}. The energy point dependent correction factors for $\mu^+$ identification efficiencies are $(87.6{\text -}93.9)\%$ with uncertainties of $(0.4{\text -}1.1)\%$, depending on the data taking status. These efficiencies are estimated using $e^+e^-\to\gamma\mu^+\mu^-$ samples and reweighted by
the $\mu^+$ two dimensional distribution in $|\cos\theta_{\mu^+}|$ and $p_{\mu^+}$ of $D_s^+\to\mu^+\nu_\mu$.

The correction factor $f_{\rm tag~bias}^{\rm cor}$ takes into account the differences of the ST efficiencies in the inclusive and signal MC samples due to different track multiplicities. This may cause incomplete cancellation of the uncertainties of the ST efficiencies. It is estimated to be about 99.57\% after considering the differences of the efficiencies of tracking or PID of $K^\pm$ and $\pi^\pm$, as well as the selections of neutral particles between data and MC simulation.

In this analysis, the shapes of signal candidates are divided into two types: one describes the signal candidates with transition $\gamma(\pi^0)$ matched and another describes the signal candidates unmatched.
The matched type requires the angle between the flight
direction of the reconstructed $\gamma(\pi^0)$ and that of MC truth
to be less than $20^\circ$ in an event, otherwise the event is classified as the unmatched type.
Both event types have real $D^+_s\to \mu^+\nu_\mu$, but the matched type forms the peak and the unmatched type forms the combinatorial shape in the $M_{\rm miss}^2$ distribution. The fraction of the un-matched $\gamma(\pi^0)D_s^+\to\mu^+\nu_{\mu}$ is within (9-10)\% for each energy point. The average signal efficiencies for finding $\gamma(\pi^0)\mu^+\nu_\mu$ at various energy points are shown in Table~\ref{tab:sigeff}.

The background includes two components. One is the events with wrongly tagged $D_s^-$ decays (``non-$D_s^-$ background''), and another is with correctly tagged $D_s^-$ decays but incorporating particle mis-identifications (``real-$D_s^-$ background'')
which is mainly from $D_s^+\to \tau^+(\to \pi^+\bar\nu_\tau)\nu_\tau$ decay.
Studies on the inclusive MC sample show
that the two components make
comparable contributions and do not peak in the signal region.
The total background fraction is about $8.2\%$.

\section{Branching Fraction Measurement}

To obtain the BF of $D_s^+\to\mu^+\nu_\mu$,
we perform a simultaneous fit to the $M_{\rm miss}^2$ distributions at
the eight energy points. The BF of $D_s^+\to\mu^+\nu_\mu$ at different energy points
is the common parameter in the fit.
The signal shapes with transition $\gamma(\pi^0)$ matched and unmatched are described
by individual simulated shapes derived from the signal MC sample, where the former ones
are further convolved with a Gaussian function with free parameters to
consider the resolution difference
between data and MC simulation.
The shapes of other backgrounds are derived from the
individual MC simulated shapes.
In the fit, the BF, the ratios between the signal
events with transition $\gamma(\pi^0)$ matched and unmatched,
and the background yields are all free.

\begin{figure}[htbp]
  \centering
  \includegraphics[width=0.5\textwidth]{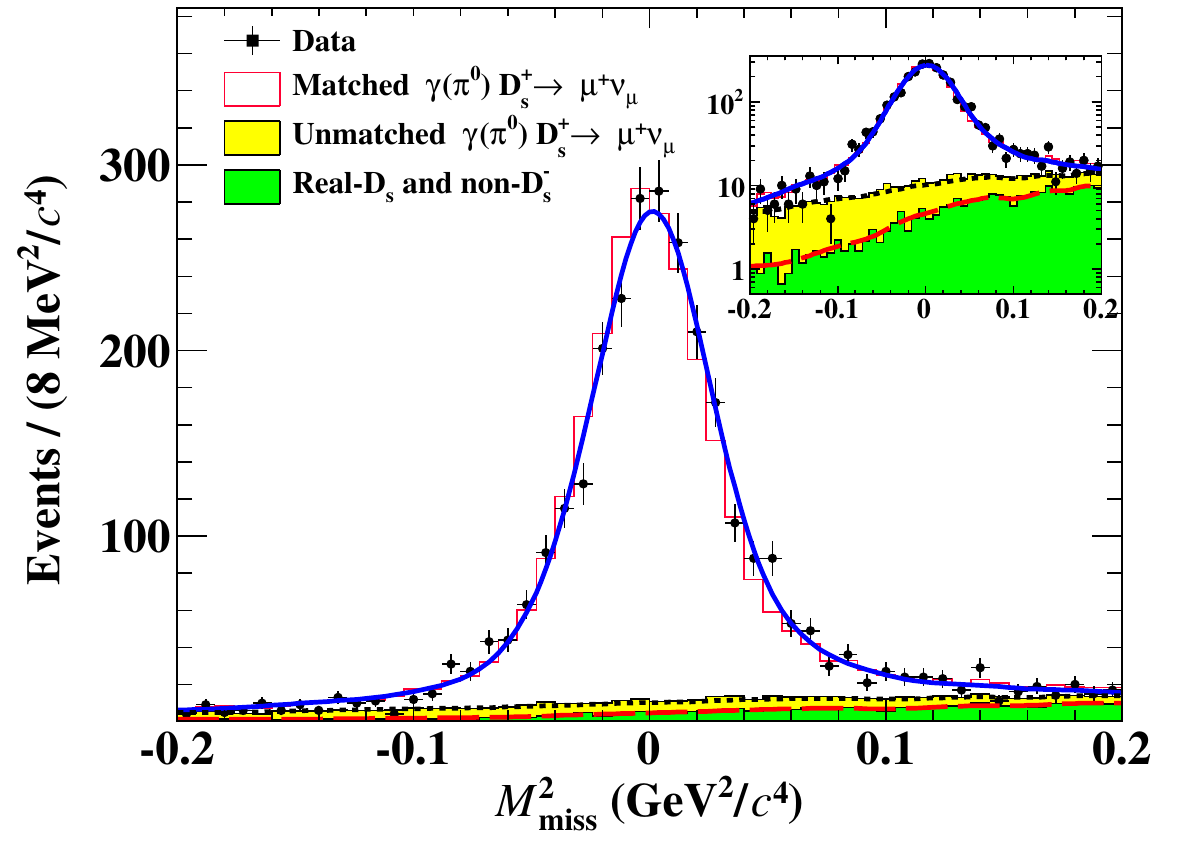}
\caption{
Fit to the $M_{\rm miss}^{2}$ distribution of the accepted candidates for $D^+_s\to \mu^+\nu_\mu$  in data.
The inset plot shows the same distribution in a log scale.
The points with error bars are data, the blue solid curve shows the best fit, and the red dashed curve shows the fitted combinatorial background shape.
Events between the red dashed and black dotted curves (yellow filled histogram) are from signals with transition $\gamma(\pi^0)$ unmatched. Events between the  black dashed and the pink histogram are from signals with transition $\gamma(\pi^0)$ matched.
The green filled histogram is the combined real-$D_s^-$ and non-$D_s^-$ background derived from the inclusive MC sample after normalization. }
  \label{fig:mm2ft}
\end{figure}

Figure \ref{fig:mm2ft} shows the fit result with all datasets.
The BF of $D_s^+\to\mu^+\nu_\mu$ is obtained to be
\begin{equation}
{\mathcal{B}}_{D_s^+\to\mu^+\nu_\mu}=(0.5294\pm0.0108)\%,
\nonumber
\end{equation}
which corresponds to a signal yield of $2514.5\pm51.6$ events.
Here, the uncertainty is statistical only.

\begin{figure}[htbp]
  \centering
  \includegraphics[width=0.5\textwidth]{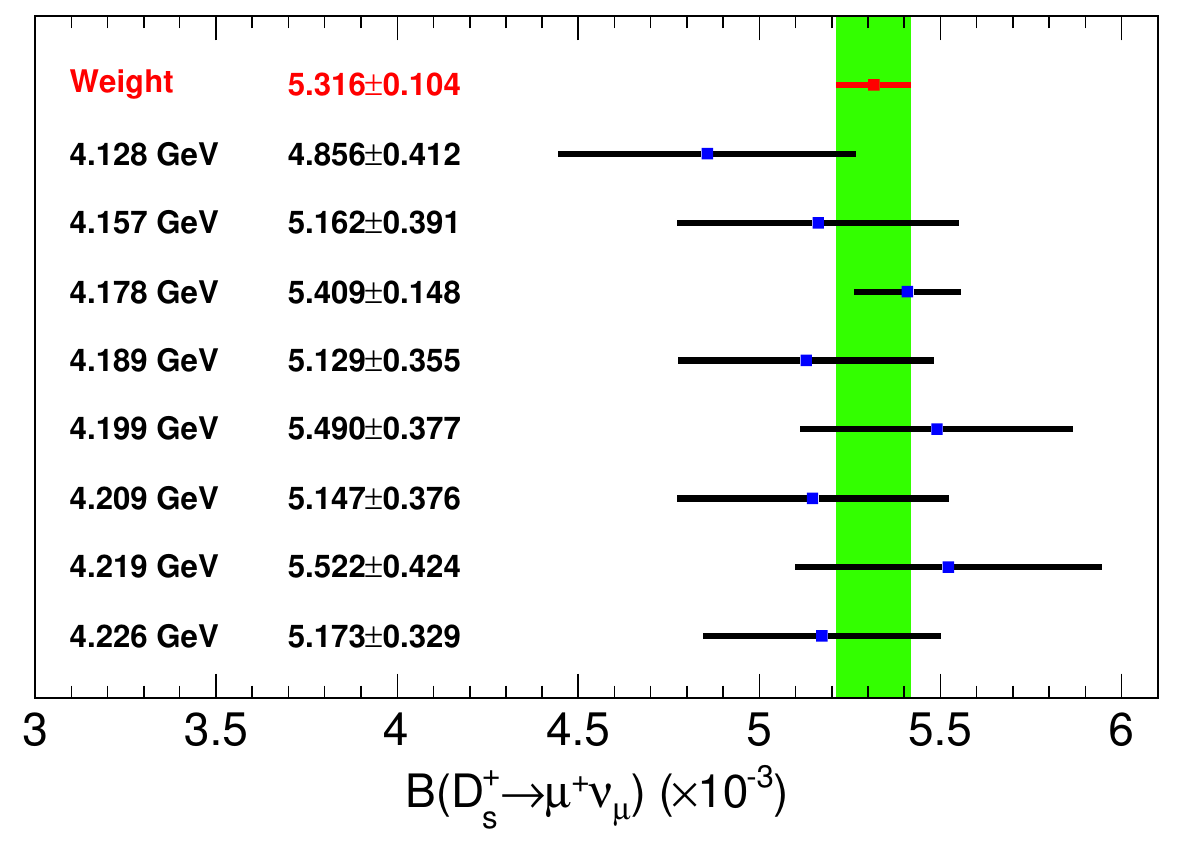}
  \caption{The obtained BFs based on individual data sets, where the shown uncertainties are statistical only. }
  \label{fig:6_dataset}
\end{figure}

As a cross check, we measure the BF of $D_s^+\to\mu^+\nu_\mu$ based on individual data sets separately and the obtained results are shown in Fig.~\ref{fig:6_dataset}. The weighted BF of $D_s^+\to\mu^+\nu_\mu$, $(0.5316\pm0.0104)\%$, is consistent with our nominal result. The different mean values and statistical uncertainties of the BF of $D_s^+\to\mu^+\nu_\mu$ are due to slightly different signal and background shapes in  different fit strategies, which has been included in the systematic uncertainty of $M^2_{\rm miss}$ fit.

\section{Systematic uncertainties}

Sources of the systematic uncertainties in the BF measurement are summarized in Table~\ref{tab:sys_crs}.
Each of them is estimated relative to the measured BF and described below.

\begin{itemize}
\item
ST yield:
The systematic uncertainty is estimated by varying the signal and background shapes in the fit.
The alternative signal shapes are obtained by varying the nominal matched angle by $\pm 5^\circ$. The relative difference of the ST yields between data and the inclusive MC sample is assigned as the systematic uncertainty. In addition, the uncertainty due to the background fluctuation in the ST yield is also considered as a systematic uncertainty. Adding these three systematic effects quadratically gives a total systematic uncertainty of 0.44\%.
\item
$\mu^+$ tracking and PID:
The $\mu^+$ tracking and PID efficiencies are studied with the control sample $e^+e^-\to\gamma\mu^+\mu^-$.
After correcting the detection efficiency by $f^{\rm cor}_{\mu\,\rm PID}$, we assign 0.24\% and 0.19\%
as the uncertainties in $\mu^+$ tracking and PID efficiencies, respectively.
\item
Transition $\gamma(\pi^0)$ reconstruction: The selection efficiencies of $\gamma$ and $\pi^0$ are studied with $J/\psi\to\pi^+\pi^-\pi^0$ decays~\cite{sys_gamma_llc}. The systematic uncertainty  is assigned to be 1\%.
\item
The least $|\Delta E|$ selection: The systematic uncertainty of selecting the transition $\gamma(\pi^0)$with the least $|\Delta E|$ method is estimated by using the control samples of $D^+_s\to K^+K^-\pi^+$ and $D^+_s\to \eta\pi^0\pi^+$. The difference of the efficiencies of selecting the transition $\gamma\,(\pi^0)$ candidates between data and MC simulation, 0.70\%, is taken as the corresponding systematic uncertainty.
\item
$E_{\rm max}^{{\rm extra~}\gamma}$ \& $N_{\rm ncharged}^{\rm extra}$ requirements: The efficiency for the requirements of $E^{\mathrm{extra}~\gamma}_{\rm max}$ and
no extra good charged track is studied with the control samples of $D^+_s\to K^+K^-\pi^+$ and $D^+_s\to K_S^0K^+$.
The systematic uncertainty is taken to
be 0.29\% considering the efficiency differences between data and MC simulation.
\item
$M_{\rm miss}^2$ fit:
The systematic uncertainty due to the signal shape with transition $\gamma(\pi^0)$ matched is estimated with an alternative signal shape of a double Gaussian function.  The systematic uncertainty due to the signal shape with transition $\gamma(\pi^0)$ un-matched is estimated by replacing the nominal shape with a second order Chebychev function. The systematic uncertainty due to the real-$D^-_s$ background is estimated by varying the weights of various background sources within $\pm 1\sigma$ of individual BFs. The systematic uncertainty due to the non-$D^-_s$ background is estimated by varying the shape smoothness parameter from 3 to 2.
For different sources, the changes of the fitted signal yield, 0.64\%, 0.18\%, 0.24\% and 0.10\%, are taken as individual uncertainties.
The total systematic uncertainty due to the $M_{\rm miss}^2$ fit is obtained to be 0.72\% by adding all four uncertainties in quadrature.

\item
Quoted BFs: The uncertainty due to the quoted BFs of $D_s^{*-}$ subdecays from the PDG~\cite{PDG2022} is examined by varying each subdecay BF by $\pm 1\sigma$. The change of the signal efficiency, 0.34\%, is taken as the associated uncertainty.
\item
Contribution from $D^+_s\to\gamma\mu^+\nu_\mu$:
The systematic uncertainty due to the contribution from the background of $D^+_s\to\gamma\mu^+\nu_\mu$ is estimated with the known upper limit on the BF of $D^+_s\to\gamma e^+\nu_e$~\cite{PDG2022}. After fixing this background yield in the $M_{\rm miss}^2$ fit, the change of the measured BF to the nominal one, 0.30\%, is taken as the corresponding systematic uncertainty.
\end{itemize}

Assuming all systematic uncertainties are independent, the total systematic uncertainty in the measurement of the BF of $D_s^+\to\mu^+\nu_\mu$ is
1.61\% by adding them in quadrature.
Here, 1.61\% corresponds to the absolute systematic uncertainty of 0.0085\% for the measured BF.

\begin{table}[htbp]\centering
  \caption{Relative systematic uncertainties in the measurement of the BF of $D_s^+\to\mu^+\nu_\mu$.}
  \label{tab:sys_crs}
  \begin{tabular}{lc}
  \hline\hline
  Source & Uncertainty (\%)\\ \hline
  ST yield        & 0.44\\
  $\mu^+$ tracking & 0.24\\
  $\mu^+$ PID      & 0.19\\
  Transition $\gamma(\pi^0)$ reconstruction & 1.00\\
  Least $|\Delta E|$ selection & 0.70\\
  $E_{\rm max}^{{\rm extra~}\gamma}$ \& $N_{\rm ncharged}^{\rm extra}$ requirements & 0.29\\
  $M_{\rm miss}^{2}$ fit & 0.72\\
  Quoted BFs & 0.34\\
  Contribution from $D^+_s\to\gamma\mu^+\nu_\mu$ & 0.30\\
  \hline
  Total & 1.61\\
  \hline\hline
  \end{tabular}
\end{table}

\section{Results of $f_{D_s^+}$ and $|V_{cs}|$}

Combining the measured BF with
the world average values of $G_F$, $m_\mu$, $m_{D^+_s}$ and
the $D_s^+$ lifetime ~\cite{PDG2022} in Eq. (1) yields
\begin{equation}
f_{D_s^+}|V_{cs}|=241.8\pm2.5_{\rm stat}\pm2.2_{\rm syst}~\mathrm{MeV}.
\nonumber
\end{equation}
Here the systematic uncertainty arises mainly from the uncertainties in the measured
BF (0.8\%) and the lifetime of the $D^+_s$ (0.4\%).
Taking the CKM matrix element $|V_{cs}|=0.97349\pm0.00016$ from the global SM fit
~\cite{PDG2022} or the averaged decay constant
$f_{D_s^+}=249.9\pm0.5~\mathrm{MeV}$ from recent LQCD calculations~\cite{FLab2018,etm2015}
as input, we determine
\begin{equation}f_{D_s^+}=248.4\pm2.5_{\rm stat}\pm2.2_{\rm syst}~\mathrm{MeV}
\nonumber
\end{equation}
and
\begin{equation}
|V_{cs}|=0.968\pm0.010_{\rm stat}\pm0.009_{\rm syst}.
\nonumber
\end{equation}
The additional systematic uncertainties according to the input parameters are
negligible for $|V_{cs}|$ and 0.2\% for $f_{D_s^+}$.

Using the BESIII combined result of ${\mathcal{B}}_{D_s^+\to\tau^+\nu_\tau}=(5.32\pm0.07\pm0.07)\%$ ~\cite{bes35}, we obtain $\mathcal{B}_{D_s^+\to\tau^+\nu_\tau}/\mathcal{B}_{D_s^+\to\mu^+\nu_\mu}=10.05\pm0.35$, where the statistical and systematic uncertainties have been combined in quadrature.
This ratio agrees with the SM predicted value of 9.75 within uncertainty, which indicates that no $\tau$-$\mu$ LFU violation is observed in $D_s^+$ sector.

\begin{table*}[htbp]\centering
  \caption{Comparisons between ${\mathcal B} (D^+_s \to \mu^+\nu_\mu)$ and  $f_{D^+_s}|V_{cs}|$ measured by various experiments. The first uncertainties are statistical, the second are systematic, and the third are from the input lifetime of $D^+_s$. The superscripts $a$ and $b$ denote the measurements are made with and without using the $d_{\mu^+}$ requirements.}
  \label{tab:bf_muv}
  \scalebox{0.8}{
  \begin{tabular}{lc r@{\,}c@{\,}l c r@{\,}c@{\,}c@{\,}c@{\,}c@{\,}c@{\,}l}
  \hline
  \hline
  Experiment &$E_{\rm cm}$ (GeV) &\multicolumn{3}{c}{Reaction chain} &\multicolumn{1}{c}{$\mathcal B$ (\%)}&\multicolumn{7}{c}{$f_{D^+_s}|V_{cs}|$ (MeV)}\\ \hline
  CLEO~\cite{cleo2009}        &4.170       &$e^+e^-$&$\to$&$D_s^{\pm}D_s^{*\mp}$ &$0.565\pm0.045\pm0.017$  &$249.8$&$\pm$&$10.0$&$\pm$&$3.8$&$\pm$&$1.0$\\
  BaBar~\cite{babar2010}      &10.56       &$e^+e^-$&$\to$&$DKX\gamma D_s^-$     &$0.602\pm0.038\pm0.034$  &$257.8$&$\pm$&$8.2$&$\pm$&$7.3$&$\pm$&$1.0$\\
  Belle~\cite{belle2013}      &10.56       &$e^+e^-$&$\to$&$DKX\gamma D_s^-$     &$0.531\pm0.028\pm0.020$  &$242.2$&$\pm$&$6.4$&$\pm$&$4.7$&$\pm$&$1.0$\\
  BESIII$^b$~\cite{bes2016}       &4.009       &$e^+e^-$&$\to$&$D_s^{+}D_s^{-}$ &$0.517\pm0.075\pm0.021$  &$238.9$&$\pm$&$17.5$&$\pm$&$4.9$&$\pm$&$0.9$\\
  BESIII$^a$~\cite{besiii2}   &4.178       &$e^+e^-$&$\to$&$D_s^{\pm}D_s^{*\mp}$ &$0.549\pm0.016\pm0.015$  &$246.2$&$\pm$&$3.6$&$\pm$&$3.4$&$\pm$&$1.0$\\
  BESIII$^b$~\cite{Hajime}    &4.178-4.226 &$e^+e^-$&$\to$&$D_s^{\pm}D_s^{*\mp}$ &$0.535\pm0.013\pm0.016$  &$243.1$&$\pm$&$3.0$&$\pm$&$3.6$&$\pm$&$1.0$\\
{\bf This work$^a$}    &{\bf 4.128-4.226}  &{\boldmath$e^+e^-$}&{\boldmath$\to$}&{\boldmath$D_s^{\pm}D_s^{*\mp}$} &{\boldmath$0.5294\pm0.0108\pm0.0085$}
&{\boldmath$241.8$}&{\boldmath$\pm$}&{\boldmath$2.5$}&{\boldmath$\pm$}&{\boldmath$2.2$}&{\boldmath$\pm$}&{\boldmath$1.0$}\\
   \hline
  \hline
  \end{tabular}
  }
\end{table*}

\section{Summary}
In summary, by analyzing 7.33~fb$^{-1}$ of $e^+e^-$ collision data
collected at $E_{\rm cm}$ from 4.128 to 4.226\,GeV with the BESIII detector,
we measure the $\mathcal{B}(D^+_s\to\mu^+\nu_\mu)$, the product of $f_{D^+_s}|V_{cs}|$, the decay constant $f_{D_s^+}$, and the CKM matrix element $|V_{cs}|$. Comparisons between the $\mathcal{B}(D^+_s\to\mu^+\nu_\mu)$ and $f_{D^+_s}|V_{cs}|$ obtained in this work and the previous measurements are shown in Table~\ref{tab:bf_muv}. These results are important to calibrate various theoretical calculations of $f_{D_s^+}$ and test the unitarity of the CKM matrix with better accuracy. We also test LFU with $D_s^+\to\ell^+\nu_\ell$ decays, and no $\tau$-$\mu$ LFU violation is found.

\section{Acknowledgement}
The BESIII Collaboration thanks the staff of BEPCII and the IHEP computing center for their strong support. This work is supported in part by National Key R\&D Program of China under Contracts Nos. 2020YFA0406400, 2020YFA0406300; National Natural Science Foundation of China (NSFC) under Contracts Nos. 11635010, 11735014, 11835012, 11935015, 11935016, 11935018, 11961141012, 12022510, 12025502, 12035009, 12035013, 12061131003, 12192260, 12192261, 12192262, 12192263, 12192264, 12192265, 12221005, 12225509, 12235017; the Chinese Academy of Sciences (CAS) Large-Scale Scientific Facility Program; the CAS Center for Excellence in Particle Physics (CCEPP); CAS Key Research Program of Frontier Sciences under Contracts Nos. QYZDJ-SSW-SLH003, QYZDJ-SSW-SLH040; 100 Talents Program of CAS; The Institute of Nuclear and Particle Physics (INPAC) and Shanghai Key Laboratory for Particle Physics and Cosmology; ERC under Contract No. 758462; European Union's Horizon 2020 research and innovation programme under Marie Sklodowska-Curie grant agreement under Contract No. 894790; German Research Foundation DFG under Contracts Nos. 443159800, 455635585, Collaborative Research Center CRC 1044, FOR5327, GRK 2149; Istituto Nazionale di Fisica Nucleare, Italy; Ministry of Development of Turkey under Contract No. DPT2006K-120470; National Research Foundation of Korea under Contract No. NRF-2022R1A2C1092335; National Science and Technology fund of Mongolia; National Science Research and Innovation Fund (NSRF) via the Program Management Unit for Human Resources \& Institutional Development, Research and Innovation of Thailand under Contract No. B16F640076; Polish National Science Centre under Contract No. 2019/35/O/ST2/02907; The Swedish Research Council; U. S. Department of Energy under Contract No. DE-FG02-05ER41374.

\bibliography{bibliography}

\end{document}